\documentclass[fleqn,twoside]{article}
\usepackage[headings]{espcrc2}
\usepackage{epsfig}
\usepackage{graphicx}
\readRCS
$Id: espcrc2.tex,v 1.2 2004/02/24 11:22:11 spepping Exp $
\newcommand{\eps}{\epsilon}
\newcommand{\be}{\begin{equation}}
\newcommand{\ee}{\end{equation}}
\newcommand{\bea}{\begin{eqnarray}}
\newcommand{\eea}{\end{eqnarray}}
\newcommand{\nn}{\nonumber}
\newcommand{\as}{\alpha_s}

\title{A numerical approach to the double real radiation part of 
$e^+e^-\to 3$ jets at NNLO}

\author{G.~Heinrich\address{Institut f\"ur Theoretische Physik, 
Universit\"at Z\"urich,
Winterthurerstrasse 190, 8057 Z\"urich, Switzerland}
        \thanks{To appear in the proceedings of the International Conference 
	``Loops and 
	Legs in Quantum Field Theory", Eisenach, Germany, April 2006.}	}

\begin{document}

\begin{abstract}
We report on the sector decomposition approach to the double real emission part 
of  $e^+e^- \to 3$\,jets at NNLO.
\end{abstract}

\maketitle

\thispagestyle{myheadings}
\markright{ZU-TH 15/06}

\section{INTRODUCTION}

Precision measurements at high energy colliders in the recent past 
led to stringent tests of the Standard Model and important bounds 
on New Physics.  
Measurements of jet rates and shape observables in $e^+e^-$ annihilation 
are of particular importance, as they allow for a precise determination of 
the strong coupling constant $\alpha_s$. 
At hadron colliders, a cross section involving  $n$ jets is proportional to 
$\alpha_s^n$ at leading order, such that an accurate knowledge of 
$\alpha_s$ will be 
important at the LHC, 
where many interesting  processes contain jets 
in the final state. 

A determination of $\alpha_s$ from 
jet rates and shape observables  is described 
e.g. in~\cite{Bethke:2006ac}, where one can see that 
for LEP measurements, the experimental uncertainty 
is smaller than the theoretical one, 
which is based on resummed next-to-leading order (NLO)
calculations. As the theoretical error is 
dominated by scale uncertainties, an NNLO calculation 
will considerably improve this situation.
A future International Linear Collider  will allow for precision
measurements at the per-mille level, which offer the possibility 
of a determination of  $\alpha_s$ with unprecedented accuracy, 
provided that  theoretical predictions at NNLO are available.

\section{COMPUTATIONAL METHODS}
The calculation of $e^+e^-\to  3$\,jets at order $\as^3$ 
requires the calculation of virtual two-loop corrections 
combined with a $1\to 3$ parton phase space, 
one-loop corrections combined with a $1\to 4$ parton phase space
where one parton can become soft and/or collinear   
(``unresolved"), 
and the tree level matrix element
squared for $1\to 5$ partons where up to two partons 
can become  unresolved. 
The unresolved particles lead to a complicated infrared 
singularity structure which manifests itself 
in $1/\eps$ poles  upon phase space integration. 
These singularities have to be subtracted and cancelled with the ones 
from the virtual contributions before a Monte Carlo program 
can be constructed. 
To achieve this task, 
two different approaches have been followed, 
one relying on the manual construction of an analytical subtraction 
scheme\,\cite{Kosower:2002su,Weinzierl:2003fx,Gehrmann-DeRidder:2004tv,Kilgore:2004ty,Frixione:2004is,Somogyi:2005xz,Gehrmann-DeRidder:2005cm}, 
the other one relying on sector 
decomposition\,\cite{Hepp:1966eg,Roth:1996pd,Binoth:2000ps,Heinrich:2002rc,Gehrmann-DeRidder:2003bm,Anastasiou:2003gr,Binoth:2004jv,Anastasiou:2004qd,Heinrich:2004jv,Anastasiou:2005qj,Anastasiou:2005pn,Melnikov:2006di,Heinrich:2006sw}.
The main features of the methods based on the explicit construction  of a 
subtraction scheme are the following: 
The subtraction terms are integrated analytically  over the 
unresolved phase space,  
such that the pole coefficients are obtained in analytic form. 
This requires appropriate phase space factorisation and subtraction 
terms which are simple enough to be integrated 
analytically in $D=4-2\eps$ dimensions.
This method naturally leads to 
a close to minimal number of subtraction terms, and
 allows insights into the infrared structure of QCD.

In the sector decomposition approach, the poles are isolated 
by an automated algebraic procedure acting in parameter space,
 and the pole coefficients are integrated 
numerically. The advantages of this approach reside in the fact that 
the extraction of the infrared poles is done by the computer, 
and that the subtraction terms can be very 
complicated as they are integrated only numerically.  On the other
hand, the algorithm which isolates the poles 
increases the number of original functions, 
%and in general does not lead to the 
%minimal number of subtraction terms, 
thus producing rather large expressions. 

The application of sector decomposition 
to real radiation at NNLO  
first has been  presented  in \cite{Heinrich:2002rc,Gehrmann-DeRidder:2003bm}, 
and the combination of the sector decomposition approach with a 
measurement function  first has been proposed
in \cite{Anastasiou:2003gr}. 
A number of NNLO results based on this method have been obtained meanwhile 
\cite{Gehrmann-DeRidder:2003bm,Anastasiou:2003gr,Binoth:2004jv,Anastasiou:2004qd,Heinrich:2004jv,Anastasiou:2005qj,Anastasiou:2005pn,Melnikov:2006di}.
Its application to the double real radiation part of 
$e^+e^-\to  3$\,jets at order $\as^3$\,\cite{Heinrich:2006sw} 
is particularly challenging due to the high number of massless 
particles in the final state, which leads to a very complicated infrared 
singularity structure.

%\vspace*{3mm}

\section{SECTOR DECOMPOSITION}
The wide range of applicability of sector decomposition goes back 
to the fact that it acts in parameter space by a simple mechanism. 
The parameters can be Feynman parameters in the case of multi-loop 
integrals, or phase space integration variables, or a combination of both. 
In the following, the working mechanism of sector decomposition
will be outlined only briefly for the example of phase space integrals, 
details can be found 
in \cite{Binoth:2000ps,Binoth:2004jv,Anastasiou:2005qj,Heinrich:2006sw}.

The phase space integral of a matrix element squared,  
combined with some measurement function ${\cal J}$ 
defining a physical observable,  
typically contains ``overlapping" structures like  
\begin{eqnarray}
%\int{\rm d}\Phi^{(D)}\,|{\rm ME}|^2 \,{\cal J}&\sim & 
&&\int {\rm d}s_{13}\,{\rm d}s_{23}\,\,s_{13}^{-1-\eps}\,\,
\frac{{\cal J}(s_{13},s_{23})}{ s_{13}+s_{23}} \label{ex}\\
&\sim &\int_0^1 dx\, dy\, x^{-1-\eps}\,\frac{{\cal J}(x,y)}{ x+y}\;,\nonumber
\end{eqnarray}
where only the dependence on two of the integration variables 
is shown for pedagogical simplicity. 
In order to extract the poles in $1/\eps$, the singularities for 
$x,y\to 0$ need to be factorised. Sector decomposition 
is a way to achieve the factorisation of this type of 
entangled singularities in an algorithmic way:
First  the integration region is split into sectors 
where the variables $x$ and $y$ are ordered by multiplying 
with unity in the form 
$[\underbrace{\Theta(x-y)}_{(a)}+\underbrace{\Theta(y-x)}_{(b)}]$. 
Then the integration domain  is remapped to the unit cube: 
after the substitutions $y=x\,t$ in sector (a) and 
$x=y\,t$ in sector (b), one has
\begin{eqnarray}
I&=&\int_0^1 dx\,x^{-1-\epsilon}\int_0^1 dt
\,(1+t)^{-1}{\cal J}(x,x\,t)\label{remap}\\
&+&\int_0^1 dy
\,y^{-1-\epsilon}\int_0^1 dt\,t^{-1-\epsilon}\,
(1+t)^{-1}{\cal J}(y\,t,y)\nn
\end{eqnarray}
where the singularities are now factorised. 
For more complicated functions, several iterations 
of this procedure may be necessary, but it is easily implemented 
into an automated subroutine. Once all singularities are factored out, 
they can be subtracted  using identities like
\begin{eqnarray*}
&&\int_0^1 dx\int_0^1 dy\, x^{-1-\kappa\eps} f(x,y)=\\
&&-\frac{1}{\kappa\eps}\,
\int_0^1 dy\,f(0,y)\\
&&+\int_0^1 dx\int_0^1 dy\,x^{-\kappa\eps}\,\frac{f(x,y)-f(0,y)}{x}\;,
\end{eqnarray*} 
where we recognise the form of plus distributions.
The result can subsequently be expanded in $\epsilon$,  
such that a 
Laurent series in $\eps$ is obtained, where the pole coefficients are 
sums of finite parameter integrals which can be evaluated numerically. 

For the numerical evaluation it has to be assured that 
no integrable singularities 
are crossed which spoil the numerical convergence. 
In the double real radiation part of $e^+e^-\to 3$ jets, 
singularities which are located in the interior 
of the integration region do indeed occur. 
However, it is always possible to remap them 
to endpoint singularites by a convenient variable transformation. 
After such a remapping, they are amenable to sector decomposition.

\section{APPLICATION TO $e^+e^-\to 3$ JETS AT NNLO}

In order to calculate the  double real 
radiation part of $e^+e^-\to 3$\,jets at NNLO, 
sector decomposition is 
applied to extract the poles appearing in massless $1\to 5$ 
particle integrals.
As a simple example, let us consider the 5-particle cut of the 
ladder graph 
shown in fig.~1. 
\begin{picture}(180,100)
\put(25, 40){\epsfig{file=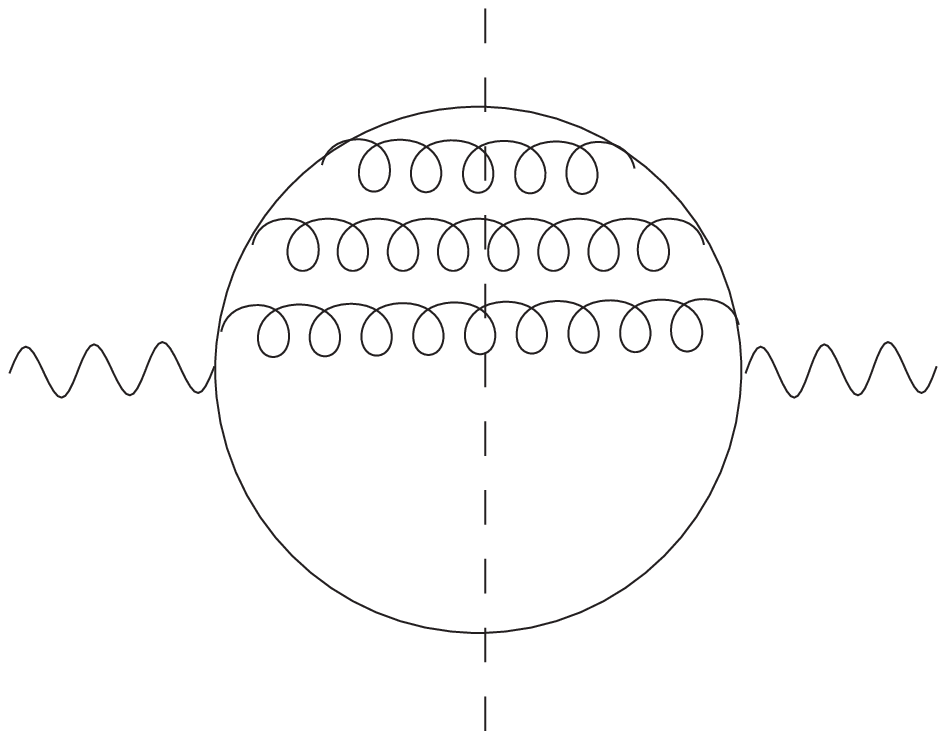,height=2.5cm}}
\put(10, 22){Figure 1. The ladder graph}
\end{picture}
\addtocounter{figure}{1}

%\begin{figure}[htb]
%\begin{center}
%\includegraphics[height=2.5cm]{C5.eps}
%\includegraphics[height=6.cm]{toprain.eps}
%\vspace*{-7mm}
%\caption{The ladder graph}
%\label{fig1}
%\end{center}
%\end{figure}

\vspace*{-6mm}

\begin{figure}[htb]
%\begin{center}
\includegraphics[height=4.8cm]{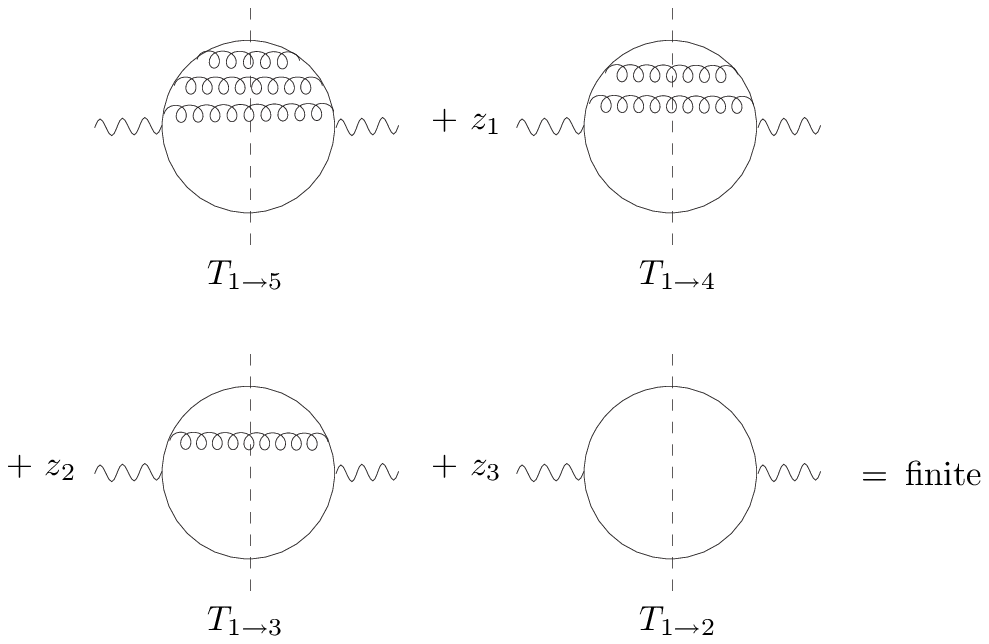}
%\end{center}
\vspace*{-7mm}
\caption{Cancellation of IR divergences in the sum over all cuts 
of the renormalised graph}
\label{fig2}
\end{figure}

Sector decomposition leads to \cite{Heinrich:2006sw}
\begin{eqnarray}
T_{1\to 5}
&=&-C_F^3\left(\frac{\as}{4\pi}\right)^3\,T_{1\to 2}
\left\{
\frac{0.16662}{\eps^3}\right.\nonumber\\
&&\left.+\frac{1}{\eps^2}\,
[1.4993-0.4999\,\log{\left(\frac{q^2}{\mu^2}\right)}]\right.\nn\\
&&\left.+\frac{1}{\eps}\,[ 5.5959-4.4978\,
\log{\left(\frac{q^2}{\mu^2}\right)}\right.\nonumber\\
&&+\left.
0.7498\,\log^2{\left(\frac{q^2}{\mu^2}\right)} ]\,+\,\mbox{finite}
\right\}\;,
\label{t5}
\end{eqnarray}
where the numerical accuracy is  better than 1\%.
%and $T_{1\to 2}$ is the leading order 
%contribution $\gamma^*\to q\bar{q}$. 
The correctness of the result can be checked by  exploiting  
%the KLN theorem\,\cite{Kinoshita:1962ur,Lee:1964is}, which implies
the fact that the sum over all cuts of a given 
(UV renormalised) diagram must be infrared finite. 
This is shown diagrammatically in fig.~\ref{fig2}.  
After UV renormalisation, 
we obtain the condition 
\begin{eqnarray}
T_{1\to5}+z_1\,T_{1\to 4}+z_2\,T_{1\to 3}+z_3\,T_{1\to 2}=\mbox{finite}\;,
\label{cfin}
\end{eqnarray}
where $T_{1\to i}$ denotes the diagram with $i$ cut lines.
%%%%%%%%%%%%%%%%%%%%%%%%%%%%%%%

\vspace*{-7mm}

\vspace*{-7mm}

The renormalisation constants $z_i$ (in Feynman gauge) are 
given by\,\cite{Binoth:2004jv,Heinrich:2006sw} 
\begin{eqnarray}
z_1&=&C_F\frac{\as}{4\pi}\,\frac{1}{\eps}\, ,\, 
z_2=C_F^2\left(\frac{\as}{4\pi}\right)^2\,
\left(\frac{1}{2\eps^2}-\frac{1}{4\eps}\right)\nonumber\\
z_3&=&C_F^3\left(\frac{\as}{4\pi}\right)^3\,
\left(\frac{1}{6\eps^3}-\frac{1}{4\eps^2}+\frac{1}{6\eps}\right)\;.
\end{eqnarray}
The expressions in eq.~(\ref{cfin}) for $i<5$ 
combine to 
%\cite{Heinrich:2006sw}
\begin{eqnarray}
&&z_1\,T_{1\to 4}+z_2\,T_{1\to 3}+z_3\,T_{1\to 2}=\nn\\
&&C_F^3\left(\frac{\as}{4\pi}\right)^3\,T_{1\to 2}
\left\{\frac{1}{6\eps^3}+\frac{1}{2\eps^2}
\,[3-\log{\left(\frac{q^2}{\mu^2}\right)}]\right.\nonumber\\
&&+\frac{1}{\eps}\,[5.61-\frac{9}{2}\log{\left(\frac{q^2}{\mu^2}\right)}
+\frac{3}{4}\log^2{\left(\frac{q^2}{\mu^2}\right)}]\nn\\
&&+\left.
\,\mbox{finite}\right\}\;.\label{c234}
\end{eqnarray}
We can see that  the poles in (\ref{c234}) 
are exactly cancelled by the 5-parton contribution (\ref{t5}) 
within the numerical precision.

\subsection{Differential results}

Although the sector decomposition approach is considered to be 
a ``numerical method", as the pole coefficients are only calculated 
numerically, 
the isolation of the poles  is an algebraic 
procedure, leading to a set of finite functions 
for each pole coefficient as well as for the finite part. 
These finite functions are written to a Fortran program
and evaluated numerically using  Monte Carlo techniques. 
%package BASES\,\cite{Kawabata:1995th}. 
In order to obtain results which are differential in a certain 
physical observable, any 
(infrared safe) measurement function can be included 
 at the level of the 
final Monte Carlo program, which means that the 
subtractions and expansions in $\eps$ do {\it not} have to be 
redone each time a different observable is considered.
Further, the measurement function does not have to be 
an analytic function, but can be a subroutine acting on the 
four-momenta of the final state particles, 
as it is typically the case for a jet clustering routine.
The reason why such a maximal flexibility is possible 
resides in the fact that the program described 
in \cite{Heinrich:2006sw} has the architecture of a 
partonic event generator. This requires to 
keep track of the mappings of the original phase space variables 
to the variables used in each sector, i.e. each 
endpoint of the iterated decompsition tree.
These mappings will be different for each sector 
(cf. the arguments of the function ${\cal J}$ in the first and second line 
of eq.~(\ref{remap})), 
but keeping this information allows to express the 
energies and angles of the final state particles in terms of the 
sector variables 
 and thus to reconstruct the fully differential 
information on the final state. 
In this procedure, function evaluations in sectors which do not pass 
the kinematical requirements imposed by the measurement function 
are unavoidable. 
Having to deal with a large number of sectors, this  can 
lead to a serious drop in efficiency of the Monte Carlo program. 
Therefore, in order to construct a program 
which produces results within a reasonable time scale,
it is crucial to keep the number of sectors low. 
This can be achieved by (a) using optimised phase space parametrisations 
for each topology,  
(b) using information on physical limits in the decomposition algorithm. 

Optimising the phase space parametrisations means the following: 
The matrix element squared can be divided into a certain number 
of topologies, which are defined by a certain set of denominators 
(which will be combinations of Mandelstam invariants, see e.g. eq.~(\ref{ex})).
In a given phase space parametrisation, some of the invariants will naturally 
be in a factorised form, but such a form cannot be achieved for {\it all} 
invariants simultaneously. An optimised parametrisation is one where 
the number of non-factorising invariants in the denominator is kept minimal.
Therefore one has to choose different parametrisations for 
each topology, resp. for each class of topologies with the same factorisation 
properties.

Using  information on physical limits exploits the fact that  
 in most cases of physical relevance, 
the measurement function is such that it would prevent certain poles 
from arising at all if it were included in the $\eps$-expansion. 
For example, poles associated with a 2-jet configuration, where 3 
of the 5 final state particles become theoretically unresolved, 
will be killed by a 3-jet measurement function. 
It would therefore be desirable to suppress the terms associated 
with such configurations already at the level of the $\eps$-expansion, 
in order to avoid an unnecessarily large number of terms associated with 
the isolation of these poles.  
%because they will be killed later by the measurement function anyway.
On the other hand, we would like to keep  the flexibility to include 
any measurement function only at the stage of the final Monte Carlo program. 
Focusing on the process  $e^+e^-\to 3$\,jets, this dilemma has been 
solved by including some ``preselection rules" 
in the $\eps$-expansion which reject configurations which will surely be 
2-jet configurations. In the example shown here, this can be achieved by introducing 
a cut parameter $y^{\rm th}$ -- which must be smaller than any possible experimental 
resolution parameter $y^{\rm cut}$ -- for the variable $s_{1345}$, as $s_{1345}\to 0$
always corresponds to a 2-jet configuration. 
In this way, one can reduce the size of the expressions considerably 
without loosing the flexibility to specify different jet algorithms 
later.
%, as long as one does not intend to calculate 2-jet production at $N^3LO$. 

\begin{figure}[htb]
\begin{center}
\includegraphics[height=5.3cm]{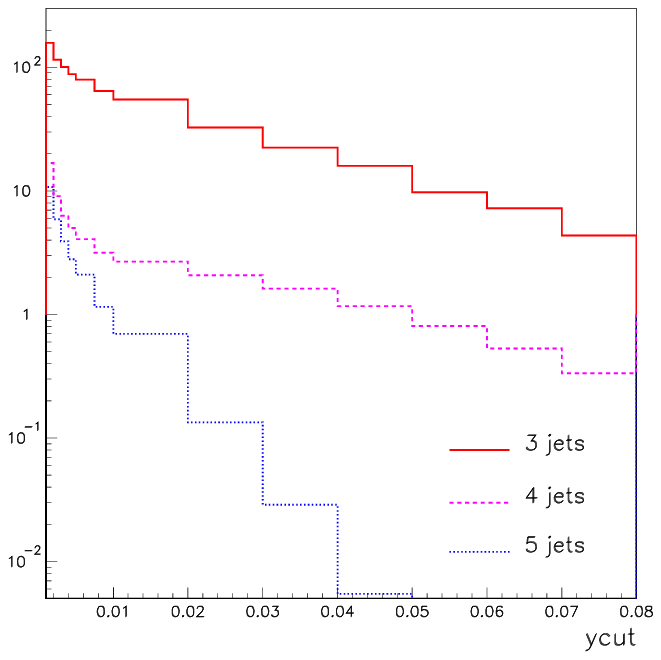}
\vspace*{-7mm}
\caption{3--, 4-- and 5--jet rates at order $\alpha_s^3$ for the toy matrix element}
\label{fig3}
\end{center}
\end{figure}

To illustrate the action of the jet function, 
3--, 4-- and 5--jet rates using the JADE algorithm\,\cite{Bethke:1988zc} 
are shown in Fig.~\ref{fig3}, based on the toy matrix element built from 
the graphs shown in Fig.~\ref{fig2}. 
%(where the 2-particle cut of course does not contribute). 

\vspace*{-7mm}

\vspace*{-10mm}

As a more complex example, the double real radiation part of a non-planar 
topology (see fig.~4) also has been calculated. 

\vspace*{9mm}

\begin{picture}(280,200)
\put(-10, 126){\epsfig{file=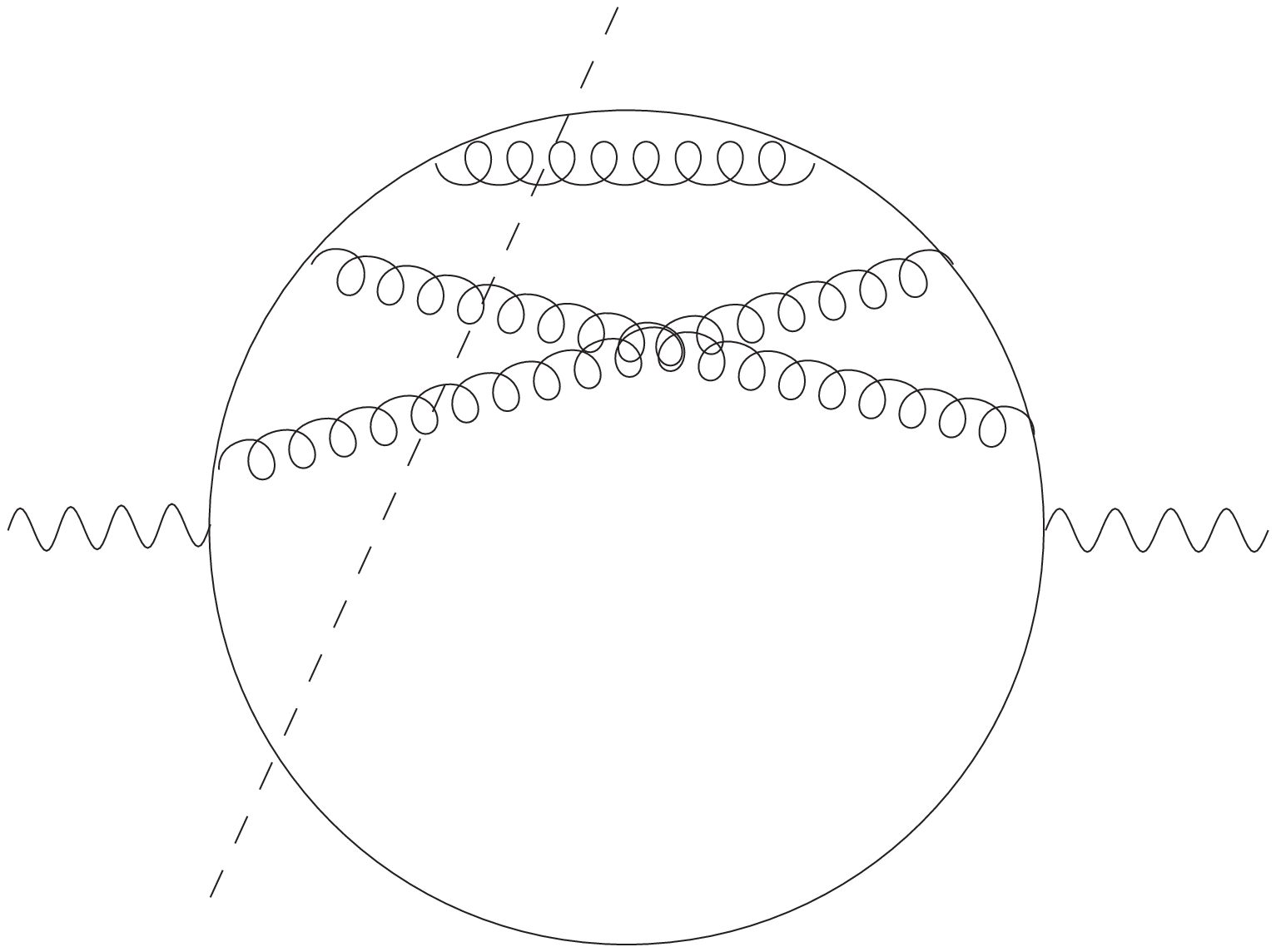,height=2.5cm}}
\put(-10, 105){Figure 4. (a) non-planar topology}
\put(95, 108){\epsfig{file=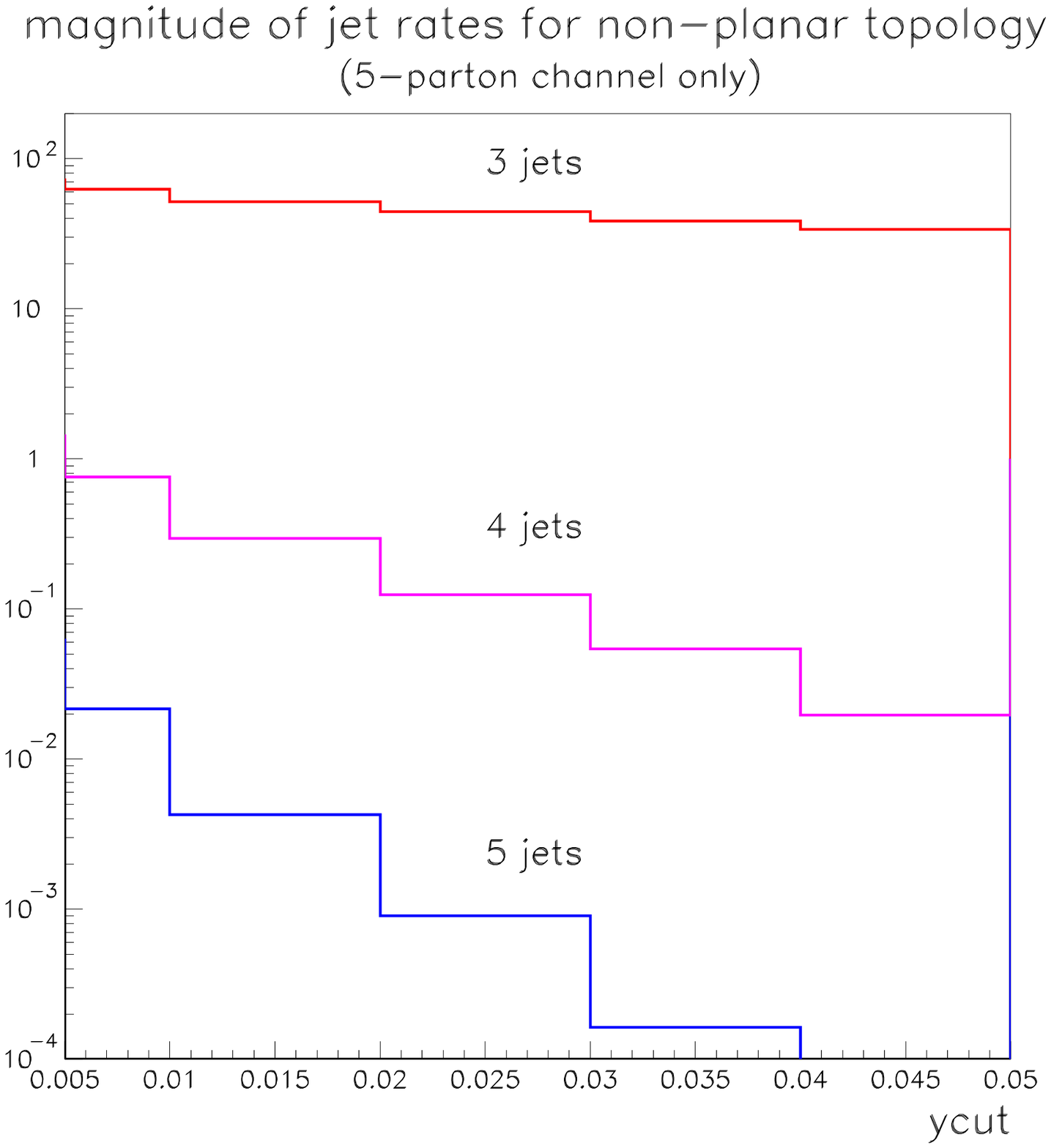,height=4.3cm}}
\put(-10, 92){(b) 3--, 4-- and 5--jet rates at order $\alpha_s^3$}
\put(3, 82){ for the non-planar topology }
\end{picture}

\vspace*{-2.6cm}

In this case, square-root terms in the 
denominator are unavoidable, which implies that the expressions produced 
by the sector decomposition will be larger due to the presence of 
more non-factorising denominators. The virtual corrections to this topology
have not yet been included. 

%\begin{figure}[htb]
%\begin{center}
%\includegraphics[height=6.cm]{cross_5parton.eps}
%\vspace*{-7mm}
%\caption{3--, 4-- and 5--jet rates at order $\alpha_s^3$ for the non-planar topology}
%\label{fig5}
%\end{center}
%\end{figure}

Note that the results shown in figs.~\ref{fig3} and 4 are unphysical, 
as they do not contain all contributions to form a gauge invariant quantity.
The purpose of these figures is merely to demonstrate the action 
of the jet function and thus the power of the method to produce differential 
results. 

The CPU time is ${\cal O}(2h)$ for a precision better than 1\%. Note that 
all topologies can be calculated in parallel, such that the  CPU time
for the full double real radiation part  will be of the order of 
the one needed for the most complicated 
topology. 

What remains to be done, besides the inclusion of the remaining topologies,  
is the combination with the 
one-loop plus single real and the two-loop virtual corrections. 
How to proceed efficiently by combining sector decomposition 
with analytical results is sketched in~\cite{Heinrich:2004jv,Heinrich:2006ku}.

\section{SUMMARY AND OUTLOOK}
The method based on sector decomposition to calculate 
$e^+e^- \to 3$\,jets at NNLO has the advantage that the isolation of the 
infrared poles is done by an automated routine and that it does not require 
the analytical integration of subtraction terms. 
However, the method produces 
large  expressions, which is an issue for a process like 
$e^+e^- \to 3$\,jets at NNLO where the matrix element 
to start with is already large. It has been described how to limit 
the proliferation of terms in the course of sector decomposition, and 
differential results have been shown for a subpart of the process.

The inclusion of massive particles within this method is 
certainly more straightforward than in analytical approaches. 
In fact, as the masses act as infrared regulators, 
the number of decompositions will be less and therefore the 
produced expressions should be smaller than in massless cases.

\vspace*{-3mm}

\section*{Acknowledgements}
\vspace*{-2mm}
This work was supported  by the Swiss National Science Foundation
(SNF) under contract number 200020-109162.

%\bibliography{bibli_ll06}

\end{document}